# Operator Responsibility for Outcomes: A Demonstration of the ResQu Model

Nir Douer
*Department of Industrial Engineering*
*Tel Aviv University*

Meirav Redlich
*Department of Industrial Engineering*
*Tel Aviv University*

Joachim Meyer
*Department of Industrial Engineering*
*Tel Aviv University*

In systems with advanced automation, human responsibility for outcomes becomes equivocal. We developed the Responsibility Quantification (ResQu) model to compute a measure of operator responsibility (Douer & Meyer, 2020) and compared it to observed and subjective levels of responsibility (Douer & Meyer, 2019). We used the model to calculate operators' objective responsibility in a common fault event in the control room in a dairy factory. We compared the results to the subjective assessments made by different functions in the diary. The capabilities of the automation greatly exceeded those of the human, and the operator should comply with the indications of the automation. Thus, the objective causal human responsibility is 0. Outside observers, such as managers, assigned much higher responsibility to the operator, possibly holding operators responsible for adverse outcomes in situations in which they rightly trusted the automation.

## INTRODUCTION

Artificial intelligence (AI) and advanced automation lead to the deployment of systems in which computers and humans share information collection and evaluation, decision-making and action implementation. In these systems, human responsibility becomes equivocal. For example, what is the human responsibility when all information arrives through a system that collects and analyzes data from various sources, without the human having any independent information? Determining human causal responsibility is particularly important when one designs or investigates intelligent systems that can lead to injury and even death, such as autonomous vehicles or automated control systems in industrial processes that involve hazardous materials.

Numerous philosophical and legal studies investigated different facets of the concept of responsibility (e.g., Vincent, 2011). When humans interact with intelligent systems, *role responsibility* refers to the specific roles and duties assigned to the humans, for which they are accountable to others. However, this role assignment does not specify the causal association between the human actions and the overall outcomes. This relation is defined by *causal responsibility*, which describes the actual human contribution to the outcomes and consequences.

Causal responsibility depends on a person's ability to control a system and the resulting consequences (Noorman & Johnson, 2014). As intelligent systems and automation become more sophisticated, users may not be sufficiently involved to be rightly considered fully responsible for the outcomes of processes (Crootof, 2015; Cummings, 2006; Docherty, et al., 2012). This leads to a "responsibility gap" in the ability to divide causal responsibility between humans and systems (Docherty et al., 2012; Johnson et al., 2014; Matthias, 2004).

Although there may be a demand for "meaningful human control" of systems (Santoni de Sio & van den Hoven, 2018), there may be cases when the human cannot supervise the system adequately, or when humans have to make decisions, based exclusively on input from automated functions they cannot evaluate independently (Pritchett, Kim & Feigh, 2014). There exist different, and at times conflicting interpretations and policies regarding meaningful human control, and system designers lack models and metrics to address this issue analytically (Canellas & Haga, 2015).

To bridge the responsibility gap, we developed a theoretical Responsibility Quantification model (the ResQu model) of human responsibility in intelligent systems (Douer & Meyer, 2020). It enables us to divide causal responsibility between the human and the intelligent system. Using information theory, we defined a quantitative measure of causal responsibility, based on the characteristics of the operational environment, the system and the human, and the function allocation between them. The ResQu model defines responsibility as the expected share of the unique human contribution to the overall outcomes. This measure also quantifies meaningful human control, which requires the human to have some causal responsibility for the outcomes.

The ResQu model is a normative theoretical model that computes a rational human's optimal level of responsibility. However, people may act non-optimally when interacting with intelligent systems (e.g., Goddard, Roudsari & Wyatt, 2011; Parasuraman & Riley, 1997). In a second study, we assessed the descriptive abilities of the ResQu model in controlled laboratory interactions with an automated decision aid (Douer & Meyer, 2019). We showed that the ResQu model can serve as a descriptive model for predicting actual human responsibility or human perceptions of their own responsibility. There was also a systematic tendency to attribute more responsibility to another person than to oneself, in the same situation, in a manner that resembles the "fundamental attribution error" in social psychology (Andrews, 2001; Ross, 2018). In a third laboratory study, we explored differences between users' and observers' subjective assessments of responsibility. We showed that actors tended to relate adverse outcomes more to system characteristics than to their own limitations, while observers failed to consider the contribution of the system to adverse outcomes, when evaluating actors' responsibility (Douer & Meyer, in press)

In this paper, we apply the ResQu model to a real-world setting, namely fault scenarios in the control room of a cheese-making factory (dairy) in Israel. We calculated the objective ResQu responsibility measures for the control room worker in the selected scenario and compared them to subjective assessments of the operator's responsibility, made by different functions in the diary.

## PRODUCTION PROCESS AND FAULT SCENARIO

The plant produces cheeses of various kinds, 24 hours a day, every day of the week, except for Saturdays. The production line includes a milk pasteurization system, a pipe system that transports milk and milk products between different production stations, 30 large containers for the storage of incoming milk, other tanks for storing different substances (milk sub-products, or cleaning materials), a large number of automatic packaging machines, and cooling systems. Also installed is a cleaning system that is connected to each of the milk storage containers. It is used to wash and clean the containers and the pipe system when a container is emptied from milk, before new milk can be pumped into the container.

All processes in the production and storage systems are controlled from a single control room, run 24-hours a day by an operator and a shift supervisor, in two 12-hour shifts. Workers in the control room aim to ensure the functioning of the system and have to intervene in fault situations. To do so, they use a computerized control system, which displays the factory's piping and instrumentation diagram (P&ID), including the states of various containers, pipes, taps, milk flow sensors, temperatures, and various alerts that indicate technical faults that require the operator's intervention. In case of such an alert, the operator (with the assistance of the shift supervisor) has to select the correct response, which is then carried out by the automated control system. Selection of an incorrect action can lead to substantial losses. The duties of the operator and the shift supervisor are not limited to interactions with the automated control system. They are also responsible for manually connecting pipes between the milk delivery trucks and the milk containers, physically checking various faults when alerted by the control system, informing maintenance personnel about technical issues, and more. Thus, at times, they operate under high workload or one of them must leave the control room.

For the current study, we selected a rather common fault scenario. It is a malfunction in a container-mounted sensor, used to alert that the container is empty of milk and needs to be rinsed and cleaned. In normal operations, each container has periods during which it contains milk (to produce various dairy products) and periods when it is empty (from the time the milk runs out to the moment it is refilled with new milk). Before refilling it with new milk, the container needs to be cleaned. A container-mounted sensor activates an indication on the control room's P&ID display, stating that a container is empty. Upon such an alert, the control room operator should initiate container cleaning by pressing a button that starts an automatic flushing process with cleaning material. When this process is complete, the container can be refilled with milk. The container-mounted sensor sometimes malfunctions and is unable to detect the presence of milk, causing the automated system in the control room to falsely indicate that the container is empty. The sensor's design assures that it cannot falsely indicate that there is milk in an empty container. According to factory data, the failure scenario occurs on average once a month, in one of the 30 milk containers.

In the above fault scenario, if the operator complies with the alert and presses the automatic flushing button, detergents are pumped into the container, ruining all remaining milk. The operator's mistake is usually detected after about 5 minutes, as the cheese production line will start to release green colored milk product, which are mixed with detergents, and the production line workers will alert the control room operator to stop the milk flow from the container. On the other hand, there may be cases when the operators might suspect that the alert is false, for example, if the operators remember that they recently filled the specific container with newly arrived milk, making it unlikely that the container is empty. In this case, the operator can go to the specific container and physically examine whether it still contains milk. To do so, the operator leaves the control room for about 15 minutes, goes to the container area, climbs a ladder to the top of the container, opens the top lid and examines the milk level. If a malfunction in the container-mounted sensor is detected, the operator informs maintenance personnel, which repair the sensor.

## OBJECTIVE RESPONSIBILITY QUANTIFICATION

In the context of the above fault scenario, the automated control system serves as a binary alert system for the presence of milk in the milk container. In this case, the ResQu model is reduced to relatively simple formulas.

Let *X* denote the binary set of the action alternatives for the human, and *Y* denote the binary classification set for the system. Then, the human causal responsibility is defined as

$$Resp(X) \stackrel{\text{def}}{=} \frac{H(X/Y)}{H(X)} = \frac{H(X,Y)-H(X)}{H(X)} \tag{1}$$

where *H(X)* is Shannon's entropy, which is a measure of uncertainty related to a discrete random variable *X*

$$H(X) = -\sum_{x\in\chi} p(x)log_2 p(x) \tag{2}$$

and *H(X/Y)* is the conditional entropy, which is a measure of the remaining uncertainty about a variable *X* when a variable *Y* is known.

$$H(X/Y) = -\sum_{y\in\gamma} p(y)\sum_{x\in\chi} p(x/y)\, log_2 p(x/y) \tag{3}$$

The ratio *Resp(X)* quantifies the expected exclusive share of the human in determining the action selection variable *X*, given the system's classification *Y*. By definition, $0 \leq Resp(X) \leq 1$. $Resp(X)=1$ if, and only if, the human action selection *X* is independent from the system's classification result *Y*. $Resp(Z)=0$ if, and only if, *Y* completely determines *X*, in which case the human actions are exclusively determined by the system's classifications.

To calculate the ResQu responsibility measure for the selected fault scenario at the dairy, one must deduce the underlying distributions of *Y* and *X* and their mutual relation.

On average, the milk in a container suffices for 4 hours of production, and flushing and cleaning an empty container takes 1 hour. Thus, on average, in a 24 hour day there are 24/(4+1)=4.8 milk cycles. In a working month, the factory works for 24 hours a day on 27 days. This amounts to an average of 129.6 milk cycles for each container.

On average, there are 4.8 true alerts a day (at the end of each milk cycle) for each container, indicating that the container is empty (True Positives). A fault in one of the 30 milk container sensors occurs on average once a month (27 working days). Thus, for a single container, there is a daily average of (1/30)(1/27)=0.00123 False Alerts, which falsely indicate that the container is empty. Thus, given an alert that a certain container is empty, the probability that the alert is true is 4.8/(4.8+0.0012)=0.99975. This is the Positive Predictive Value *(PPV*) of the automated system.

On average, every 5-7 hours each container is refilled with milk, which requires the operator to manually connect pipes to the milk containers. Thus, the operator is somewhat able to assess independently whether a specific container contains milk by recalling when he or she last refilled that container. However, due to the large number of milk containers, this subjective assessment is not perfect. The operators estimated that they have a relatively good chance, of .9, to correctly and independently assess the true state of each milk container, whether it is empty or contains milk. Nevertheless, they may initiate cleaning of a container only when there is a relevant alert to do so. Table 1 summarizes the joint distribution of the true state of the container, and the operator's perception of it, upon alert arrival.

TABLE 1 - JOINT DISTRIBUTION OF THE TRUE STATE OF THE CONTAINER, AND THE OPERATOR'S PERCEPTION OF IT, UPON ALERT.

| | | Operator's perception | | Marginal distribution of container |
|---|---|---|---|---|
| | | "Empty" | "Not Empty" | |
| True state, upon alert | Empty | 0.899775 | 0.099975 | 0.99975 |
| | Not Empty | 0.000025 | 0.000225 | 0.00025 |

Using Bayes' theorem and Table I, with an alert, if the operator thinks that the container is empty, there is a 0.899775/(0.899775+0.000025 )=0.99997 probability that it is indeed empty. If the operator thinks that the container is *not* empty, there is still a very high probability of 0.099975/(0.099975+0.000225)=0.99775 that it is actually empty, due to the very high *PPV* of the system.

The operator's action selection, given an alert, depends on the costs associated with wrong decisions (correct decisions and their related costs and benefits represent the nominal state of normal operation). When there is a false alert that a container is empty, and the operator complies with the alert and initiates automatic flushing, there is a direct average cost of 5,040 Israeli Shekel (ILS), due to the price of the mean level of milk content in a milk container and the price of the detergents used. Besides this direct cost, there is also an indirect high cost, due to the damage to the operator's reputation and competence in the eyes of the factory management and production line workers (the operator is not fined personally for such a mistake). When there is a true alert, but the operator wrongly thinks that it is false and goes to physically check the milk content in the container, there is an average cost of 750 NIS, which reflects the revenue loss, due to the 15 minutes delay in starting the next production cycle. Besides this direct cost, there is also an indirect cost, due to the physical effort involved in walking to the tanks, climbing the container and examining the milk level, and the cost of leaving the control room for 15 minutes, which may be problematic when the workload in the control room is high (e.g. in the morning, when many tank trailers with fresh milk arrive).

Given the above costs and probabilities, one can compute the optimal action selection that minimizes expected costs. Assume that when the alert arrives, the operator also thinks that the container is empty. Thus, the probability that the container is indeed empty is 0.99997, so if the operator wrongly chooses to physically go and inspect the milk content, there will be an associated average cost of 0.99997·750=749.98 ILS. However, there is still a very low chance of 0.00003 that the container is not empty, so there is an associated average cost of 0.00003·5040= 0.14 ILS to initiating the cleaning process. Thus, the optimal strategy, in this case, is to comply with the alert, by initiating the cleaning process. It is easy to show that this is also the optimal action when the operator sees an alert, but thinks that the container is not empty, upon alert arrival. Thus, the optimal strategy is to always comply with the alert, even if the operator suspects that it is false. This results from the very high *PPV* of the alert system, compared to the human, and the cost structure.

The objective ResQu responsibility measure, using (1), shows that with the above optimal strategy, the automated sensor classification (denoted by *Y*) completely determines the operator's response (denoted by *X*). The operator should always initiate a cleaning process upon an alert, and do nothing when the signal is that "there is still milk". In terms of entropies this actually means that $H(X/Y)=0$ since, given *Y* (the type of systems indication), there is no uncertainty left regarding *X* (the operators' action selection). Thus, the ResQu model objective measure for human responsibility is zero. This is reasonable, since with the optimal strategy, the operator simply transforms system indications into prescribed actions, and hence does not truly contribute to the action selection process.

## SUBJECTIVE RESPONSIBILITY ASSESSMENTS

We presented descriptions of three fault scenarios, related to the failure discussed above, to different functions in the plant: a control room operator, a shift manager in the control room, a production line worker, the quality assurance manager of the dairy, and the chief financial officer of the diary. They provided their subjective judgments on the action selection and responsibility of the control room worker in each of the scenarios. The action selection was rated on a 1-5 scale, where 1 is "didn't act properly" and 5 is "acted properly" and the responsibility was rated on a 0-100% scale, where 0 is "not responsible at all for the losses" an 100% is "fully responsible for the losses". We also asked the participants to explain their choices.

**Scenario I – A large financial loss due to incorrect cleaning of a container, while still full of milk**

In this scenario, during a busy morning shift, the control room was alerted that one container is out of milk and needs to be cleaned. The shift manager (who had filled that container with milk about half an hour earlier) performed another task and was not in the control room. The operator complied immediately with the alert and initiated a cleaning process. About 5 minutes later, a production line worker rushed to the control room and updated that this was a mistake and that the production must be stopped. About 2,000 liter of milk in the container had to be discarded, which is equivalent to a loss of 10,000 ILS. Figure 1 presents the subjective scores for five different functions in the dairy for this scenario. The control room operator's assessments clearly differed from those of all the other functions in the plant.

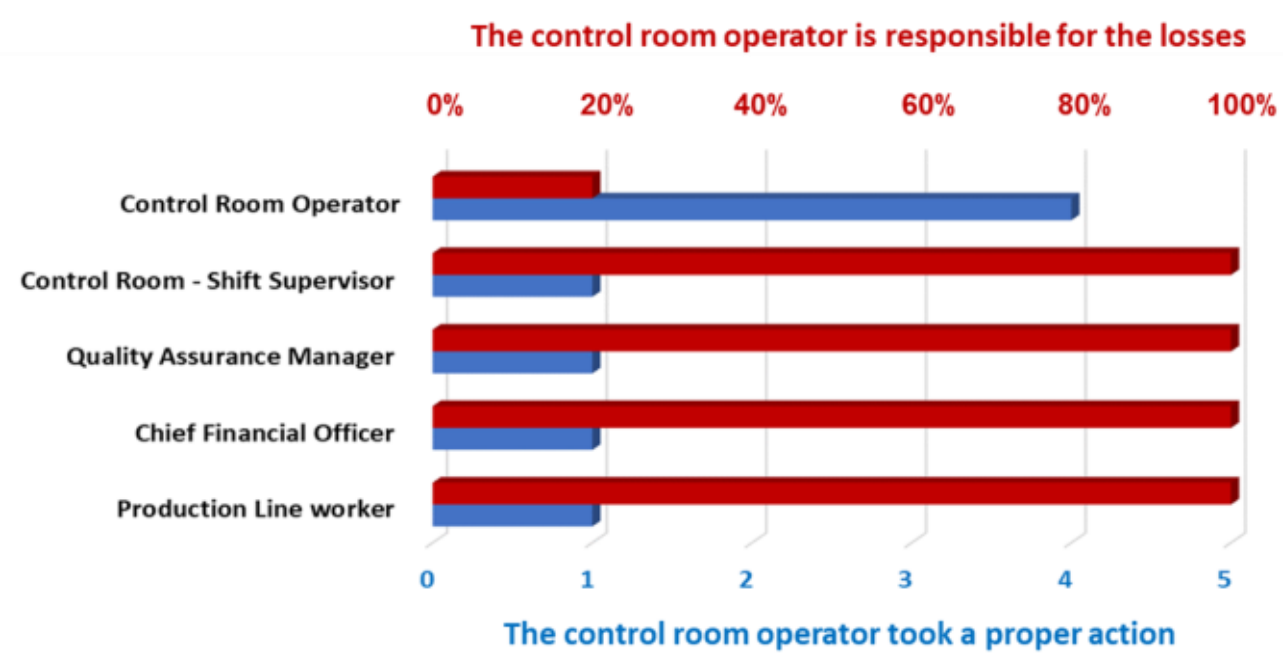

Figure 1. Subjective scores in the 1st Scenario.

The operator stated that the operator's action in the scenario was mostly correct and that the operator's responsibility for the losses is low. He explained that the operator acted in accordance with the alert, and that it is customary and logical to do so, because there are many alerts during the day and the alert system is highly reliable. The shift supervisor rated the operator's action as wrong and held the operator fully responsible for the losses. He explained that the operator could have consulted with the shift supervisor, which might have prevented the damages.

The Chief Financial Officer's (CFO) and Quality Assurance Manager's (QAM) scores were similar to those given by the shift supervisor. In their view, a full tank of milk had to be discarded because of the operator's mistake and because he did not consult with the shift manager. Thus, the operator acted wrongly and was fully responsible for the adverse outcomes. The production line worker's assessment was similar. From his point of view, in general, the control room operators are fully responsible for such mistakes, which stop production and cause more work for the production workers (e.g. extra cleaning activities, re-pasteurization, etc.).

**Scenario II – A medium financial loss due to incorrect cleaning of a partially filled container**

In this scenario, toward the end of an afternoon shift, the control room was alerted that one container is out of milk, and it needs to be cleaned. The operator remembered that this container was filled with milk some hours ago, so the alert seemed sensible. Due to being tired after a busy shift, the operator chose to comply with the alert and initiated a cleaning process. About 5 minutes later, a production line worker rushed to the control room and updated that this was a mistake and that the production must be stopped. About 400 liter of milk that were in the container had to be discarded - a loss of 2,000 ILS. This scenario is different from the previous one in two aspects. First, the incurred loss is smaller (2,000 ILS vs. 10,000 ILS). Second, there is a higher correlation between the timing of the alert and the possible state of the container, increasing the probability that the container was empty. Figure 2 presents the subjective scores for the five functions in the dairy in the second scenario.

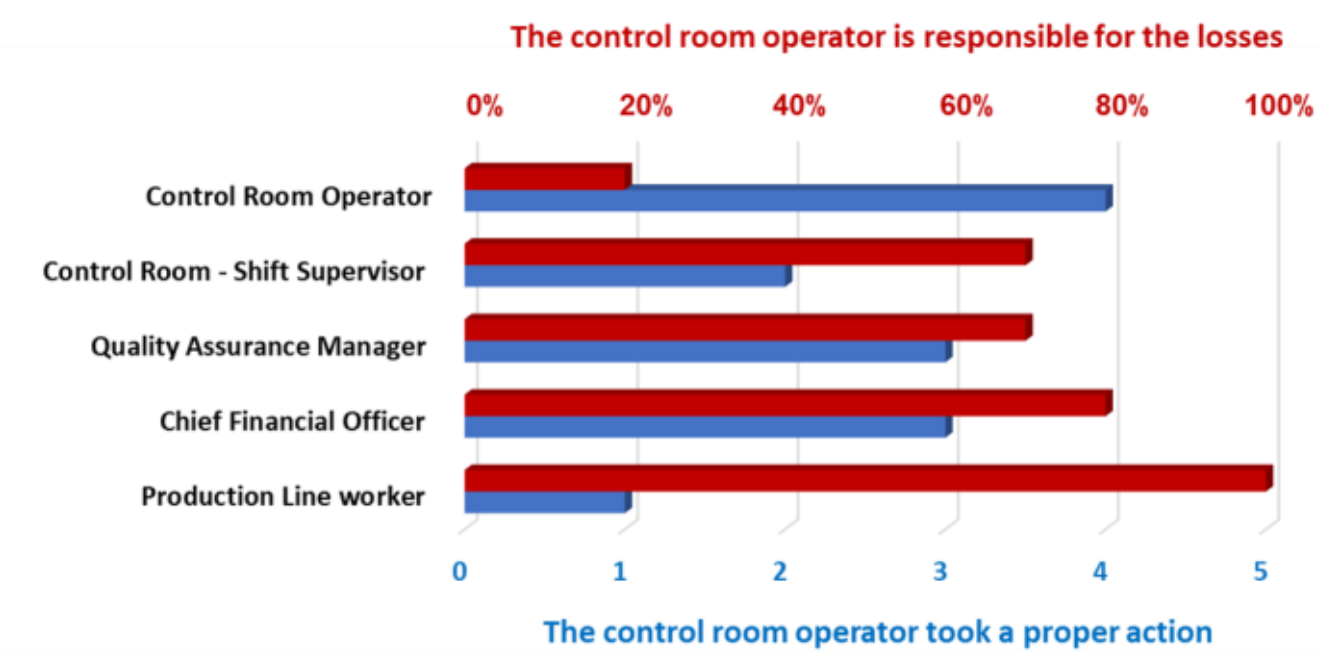

Figure 2. Subjective scores in the 2nd Scenario.

The operator's assessments were identical to those in the first scenario, justifying the action of the control room worker and assigning low responsibility for the loss. In his explanations he mentioned also the high correlation between the times since the container was filled and the timing of the alert, so the alert seemed logical. The shift supervisor, the CFO and the QAM justified the operator's action, more than in the first scenario, and attributed only partial responsibility for the losses (compared to the full responsibility they gave in the first scenario). The difference was due to the long time since the container was filled with milk. However, because there was still a considerable loss, they did find the operator responsible, though to a lesser extent than in the first scenario. The production line worker's assessments were identical to those in the first scenario, attributing full responsibility to the operator.

**Scenario III – A small financial loss due to the operator questioning the validity of the alert**

In this scenario, during a busy morning shift, only the operator is temporarily in the control room. An alert is issued that one container is out of milk and needs to be cleaned. It seemed to the operator that he had filled this container only two hours ago, and thus he suspected that this is a false alert. He chose to go to the storage area and physically check if there is still milk in the container. In the inspection, he discovered that the alert was correct and there was no milk, so the container needed to be cleaned. He went back to the control room and initiated cleaning. The inspection caused a 15 minutes delay in the production, which is equivalent to a loss of 750 ILS. Figure 3 presents the subjective scores for the five functions in the dairy in the third scenario.

In this scenario, the operator again justified the action of the control room worker, explaining that if the operator really suspected a false alert, he should conduct another examination, due to the high costs of falsely initiating cleaning. The QAM supported the operator's assessments, stating that the control room worker acted correctly and has low responsibility.

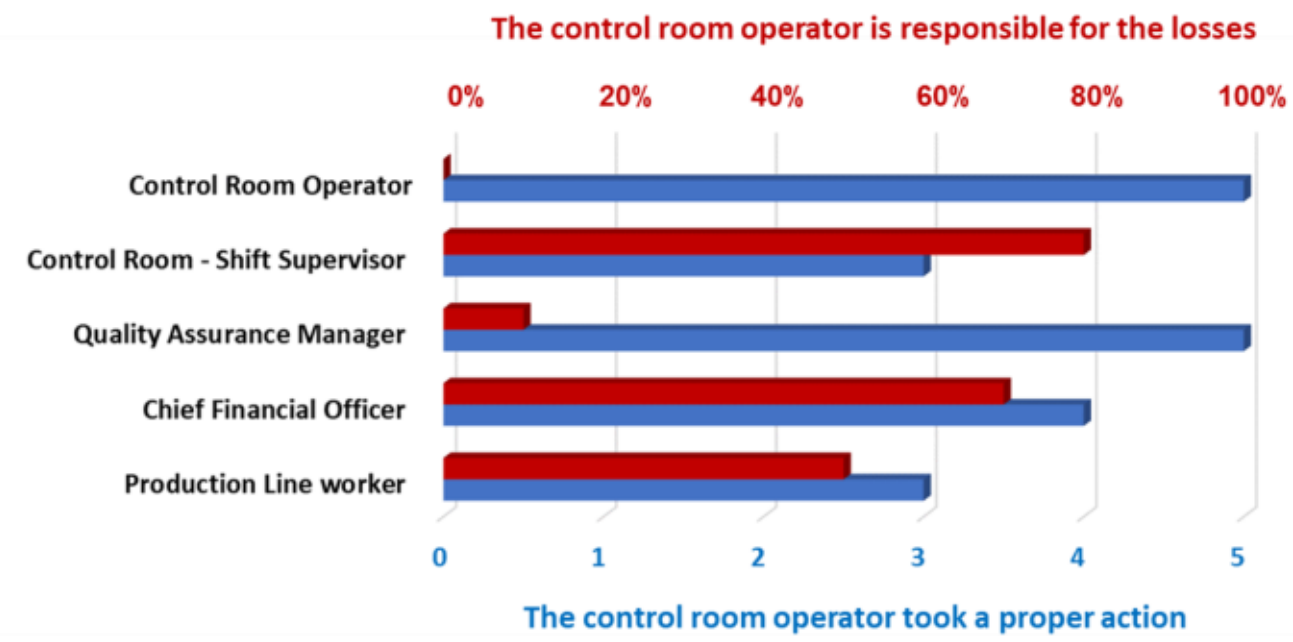

Figure 3. Subjective scores in the 3rd Scenario

She explained the she always prefers to check if there is any doubt related to the production process, to avoid unnecessary major errors, even if this involves "minor" costs. Most other evaluators also believed that the operator acted correctly, because the cost of the action was rather low, and since, if it was a false alert, the remaining milk would have to be discarded. At the same time, the shift supervisor, the CFO, and the production worker, still considered the operator responsible for the unnecessary financial loss, due to the unnecessary delay.

## DISCUSSION

In the selected fault scenarios, due to the high *PPV* of the alert system, the operator's limited knowledge about the state of the system, and the structure of the payoff matrix, a rational operator should always act according to the indications from the automation. In this case, the automation unequivocally dictates the operator's actions. The operator has no unique marginal contribution and has no causal responsibility for the outcomes.

Different functions in the plant perceived the operator's responsibility differently. Even though in the first two scenarios the operators acted optimally by fully complying with the alerts, management considered the operators to be somewhat responsible. The tendency to consider operators excessively responsible for events may cause organizations to hold operators responsible for adverse outcomes, even when they acted optimally and trusted the automation.

Our analysis is a first demonstration of the application of the ResQu model to the case of simple automation and a single type of failure scenario. This is clearly not a methodologically sound quantitative study, and the respondents' statements are anecdotal. Still, in many settings one cannot conduct systematic surveys, because there are simply too few people who fulfill specific roles in the organization. Hence research will necessarily be more qualitative in nature, similar to the study we present here. It is intended to serve as a proof-of-concept and as a demonstration of the evaluation of causal responsibility in human-automation interactions.